\documentclass[12pt]{article}
\usepackage{tabularx} 
\usepackage{amsmath, amssymb}  
\usepackage{graphicx} 

\usepackage[margin=1in]{geometry} 
\usepackage{cite} 

\usepackage{graphicx}
\usepackage{amssymb,amsthm,amsmath}
\usepackage{color}
\usepackage[shortlabels]{enumitem}
\usepackage[hidelinks]{hyperref}
\usepackage{notation}
\usepackage{comment}

\newtheorem{theorem}{Theorem}[]
\newtheorem{definition}[theorem]{Definition}
\newtheorem{example}[theorem]{Example}
\newtheorem{lemma}[theorem]{Lemma}
\newtheorem{proposition}[theorem]{Proposition}
\newtheorem{corollary}[theorem]{Corollary}

\begin{document}

\AtEndDocument{%
  \par
  \medskip
  \begin{tabular}{@{}l@{}}%
    \textsc{Pedro Baptista} \\
    \textsc{Dept. of Computer Science} \\ 
    \textsc{Universidade Federal de Minas Gerais, Brazil} \\
    \textit{E-mail address}: \texttt{pedro.baptista@dcc.ufmg.br}\\ \ \\
    
    \textsc{Gabriel Coutinho}\\
    \textsc{Dept. of Computer Science} \\ 
    \textsc{Universidade Federal de Minas Gerais, Brazil} \\
    \textit{E-mail address}: \texttt{gabriel@dcc.ufmg.br} \\ \ \\
    
    \textsc{Vitor Marques} \\
    \textsc{Dept. of Statistics} \\ 
    \textsc{Universidade Federal de Minas Gerais, Brazil} \\
    \textit{E-mail address}: \texttt{vitormarques@ufmg.br}
  \end{tabular}}
  
\title{Unexpected Averages of Mixing Matrices}
\author{Pedro Baptista, Gabriel Coutinho, Vitor Marques\footnote{vitormarques@ufmg.br --- remaining affiliations in the end of the manuscript.}}
\date{August 31, 2023}
\maketitle

\begin{abstract} 

The (standard) average mixing matrix of a continuous-time quantum walk is computed by taking the expected value of the mixing matrices of the walk under the uniform sampling distribution on the real line. In this paper we consider alternative probability distributions, either discrete or continuous, and first we show that several algebraic properties that hold for the average mixing matrix still stand for this more general setting. Then, we provide examples of graphs and choices of distributions where the average mixing matrix behaves in an unexpected way: for instance, we show that there are probability distributions for which the average mixing matrices of the paths on three or four vertices have constant entries, opening a significant line of investigation about how to use classical probability distributions to sample quantum walks and obtain desired quantum effects. We present results connecting the trace of the average mixing matrix and quantum walk properties, and we show that the Gram matrix of average states is the average mixing matrix of a certain related distribution. Throughout the text, we employ concepts of classical probability theory not usually seen in texts about quantum walks.

\end{abstract}

\begin{center}
    \textbf{Keywords}\\
    average mixing matrix; characteristic functions; quantum walks.
\end{center}

\section{Introduction}
\label{Introduction}

Let $G$ be a simple undirected graph on $n$ vertices, and $A= A(G)$ its adjacency matrix. Each vertex corresponds to a qubit, and a quantum walk is defined as the evolution of the state space under a choice of a Hamiltonian. We work with the Hamiltonian in which every edge $ij$ of the graph corresponds to the operator $\sigma^i_X\sigma^j_X + \sigma^i_Y\sigma^j_Y$. The adjacency matrix is a block of this Hamiltonian corresponding to the subspace spanned by states in which one vertex is at $\ket 1$ and the remaining at $\ket 0$. Therefore, the matrix exponential 
\begin{equation*}
    U(t) = \exp(\ii t A)
\end{equation*}
\noindent
is a unitary matrix that defines a \emph{continuous-time quantum walk} with respect to the graph.

The typical behaviour of a quantum walk can be observed by computing the average of the mixing matrices of the walk over certain time intervals (discrete or continuous), or by taking the limit of this interval to be the real line. This line of investigation was introduced more than 20 years ago \cite{aharonov2001quantum} to study properties of a quantum walk in analogy to those of classical random walks. This concept was explored for continuous-time quantum walks in hypercubes in \cite{moore2002quantum}, complete graphs in \cite{adamczak2003note} and Cayley graphs, in particular circulants, in \cite{adamczak2007non}. For all of these papers, the question of whether a quantum walk that starts at a vertex and displays an average uniform distribution in the network was thoroughly considered. The (standard) average mixing matrix of a continuous-time quantum walk is defined as the matrix whose rows display these averages, that is:
\[
	\widehat{M} = \lim_{T \to \infty} \frac{1}{T} \int_{0}^T U(t) \circ \overline{U(t)} \ \textrm{d}t. 
\]
(Note that $\circ$ stands for the entry-wise product of matrices.)

If the graph is vertex-transitive\footnote{Meaning: for all pairs of vertices, there is a symmetry of the graph that maps one to the other.}, then one row of this matrix is constant if and only if all are, and this was shown to be impossible for Cayley graphs on Abelian groups (in particular circulants) on more than two vertices in \cite{adamczak2007non}. In \cite{GodsilStateTransfer12} it was shown that $\widehat{M}$ is never constant for any graph on more than two vertices. 

In this paper, we show examples of graphs for which the average mixing matrix is constant, provided the averaging is performed over non-uniform probability distributions. As far as we are aware, this approach has not been previously considered, and our results suggest the application of classical probability theory in quantum walks can lead to interesting quantum protocols.

To further this connection, we employ some probability theory, in particular the framework of characteristic functions of distributions, to analyse in detail several known facts about the (standard) average mixing matrix under this new model. Basic algebraic properties of the (standard) average mixing matrix and its connection with the combinatorics of the underlying graph were largely explored in \cite{Godsil2013}. In this paper, we show that several of these results have a counterpart for generalized average mixing matrices. The observation that the standard average mixing matrix is the Gram matrix of average states lead to results connecting its algebraic properties and symmetries of the graph in \cite{Coutinho2018}; in this paper, we show that the generalized average mixing matrices are still related to certain average states. The diagonal entries and trace of the standard average mixing matrix were carefully analysed in \cite{GodsilGuoSinkovicAMMTrees,godsil2019diagonal}, and in \cite{godsil2021sedentary} an interesting the connection between the diagonal of $\widehat{M}$ and the stay-at-home property of the quantum walk was observed. Again, we show relations between the traces of generalized average mixing matrices and properties of the quantum walks. 

In Section~\ref{sec:2}, we show basic properties of the average mixing matrix for arbitrary sampling distributions. In Section~\ref{discrete}, we show how to achieve new behaviour already with simple discrete distributions. In Section~\ref{char}, we use the technology of characteristic functions to study the average mixing matrix under continuous distributions, finding more interesting examples. In Section~\ref{cartesian}, we use covariance to show a formula for the trace of the average mixing matrix of the Cartesian product of graphs, with an application to the study of uniform average mixing. The connection between the average mixing matrix and average states is formalized in Section~\ref{average}, and finally we finish our text with some open lines of investigation.

\section{Average mixing matrix under general densities} \label{sec:2}

We will work mainly with undirected, unweighted, finite graphs, typically denoted by $G$, usually on $n$ vertices. We use $A$ to denote its adjacency matrix: the $01$ square matrix indexed by vertices, where $1$s indicate that the corresponding vertices are neighbours. This is a symmetric matrix, and in this paper we shall use the following notation for the spectral decomposition of $A$:
\[
	A = \sum_{r = 0}^d \theta_r E_r.
\]
Here $\{\theta_r\}$ is the set of distinct (real) eigenvalues of $A$, and each $E_r$ represents the orthogonal projection onto the corresponding orthogonal eigenspaces. Recall that the trace of $E_r$ is the dimension of the eigenspace. The matrix exponential 
\begin{equation*}
    U(t) = \exp(\ii t A)
\end{equation*}
is a unitary matrix that defines a \emph{continuous-time quantum walk} in $G$. The \emph{mixing matrix} of the quantum walk is
\begin{equation}
\label{equationMof_t}
    M(t) = U(t)\circ\overline{U(t)}.
\end{equation}
The $i$th column of this matrix encodes the probability density of the quantum system at time $t$, provided the initial state was $\ket i = \otimes_{j = 1}^n \ \ket{\texttt{I}[j=i]}$.

The \emph{average mixing matrix} (of the uniform density on the real line) is given by
\[ \widehat{M} := \lim_{T \to \infty}\frac{1}{T}\int_{0}^{T}U(t)\circ \overline{U(t)} \  \mathrm{d}t = \lim_{T \to \infty}\frac{1}{T}\int_{0}^{T}M(t)\  \mathrm{d}t.
\]

Recall that (see for instance \cite{Coutinho2018, Godsil2013})
\[
	U(t) = \sum_{r = 0}^d \e^{\ii t \theta_r} E_r,
\]
and from this it is easy to see that 
\begin{equation}
\label{equationAvgMixMat}
    \widehat{M} = \sum_{r=0}^d E_{r}\circ E_{r}.
\end{equation}
Proposition~\ref{PropDefinAvgMixMat} below provides a proof of this fact as a particular case. Before going on, we recall an important property for functions of random variables.

\begin{proposition}
\label{propsExpecPSF}(see for instance \cite[Section 4.1]{schervish2014probability})
Let $R$ be a random variable and $f_{R}(x)$ its probability density function. For any real-valued function of a real variable $h$,
$$\mathbb{E}[h(R)] = \int_{-\infty}^{\infty}h(x)f_{R}(x)\dd x.$$
\end{proposition}

\subsection{$\widehat{M}$ with an arbitrary sampling distribution}
\label{SubSecNewAvgMixMat}

In this paper, we will use arbitrary probability sampling distributions to define our average mixing matrices.

\begin{definition}\label{def:amm}
Let $R$ be a random variable with probability density function $f_R(t): \mathbb{R} \xrightarrow{} [0, \infty)$. Let $G$ be a graph, and $M(t)$ the mixing matrix of the quantum walk on $G$. The \emph{average mixing matrix under $R$} is defined as
\[\widehat{M}_{R} := \mathbb{E}[M(R)] = \int_{-\infty}^{\infty} M(t)f_R(t) \ \mathrm{d}t.\]
\end{definition}

\noindent
Since $M(-t) = M(t)$ is well-defined, the domain of $f_R$ is not restricted to $\mathbb{R}^+$. This enables us to work with a wider variety of distributions, such as the Gaussian family.

\begin{definition}\label{def:uniform}
    We say that a graph $G$ on $n$ vertices admits uniform average mixing with respect to $R$ if 
    \begin{equation*}
        \widehat{M}_{R} = \frac{1}{n}J,
    \end{equation*}
    where $J$ is the all-ones matrix.
\end{definition}

In probability theory, it is common to distinguish between discrete and continuous distributions. In the discrete setting, probabilities are obtained from the point-wise evaluation of the mass function, while in the continuous case, it is done by integration of the density function. To develop a consistent theory, we would like to work with both types of distribution under the same framework. This is possible with the introduction of Dirac's delta function $\delta(t)$. For further details, see \cite[Chapter 6]{BoycePrima}.

\begin{definition}$\delta(t)$ is a function defined to have the properties
\begin{enumerate}[(i)]
	\item $\delta(t) = 0$ for $t \neq 0$.
	\item $\int_{-\infty}^{\infty}\delta(t) \dd t = 1$
\end{enumerate}
We will denote $\delta_a(t):= \delta(t - a)$.
\end{definition}

Now we can work with discrete probability distributions using probability density functions. For instance, the probability density function of a random variable $X \sim \texttt{Bernoulli}(p)$ can be written as $f_X(t) = (1-p)\ \delta_0(t) + p\ \delta_1(t)$. Due to its extensive use throughout the paper, deterministic distributions $X$ of the form $f_X(t) = \delta_a(t)$ will be named Dirac distributions.

\subsection{General formulation for the standard $\widehat{M}$}
\label{SubSecProbFormAvg}
Let $R_{T}$ be a random variable with uniform distribution over an interval of length $T$ starting at $0$, for short $R_{T} \sim \texttt{Uniform}(0, T)$. Let $f_{R_T}(t)$ be the associated probability density function, that is
\begin{align*}
    f_{R_T}(t) = \frac{1}{T}I(t)_{[0,T]},
\end{align*}
where $I(t)_{S}$ is the indicator function that is equal to $1$ whenever $t \in S$. It is straightforward to notice that we can redefine the standard average mixing matrix in terms of this probability density function

\begin{equation*}
    \widehat{M} = \lim_{T \to \infty}\int_{-\infty}^{\infty}U(t)\circ\overline{U(t)} f_{R_T}(t) \dd t.
\end{equation*}

\subsection{A formula for $\widehat{M}_R$}

Our first result displays a formula for $\widehat{M}_R$ analogous to \eqref{equationAvgMixMat}.

\begin{proposition}
\label{PropDefinAvgMixMat}
Let $R$ be a sampling distribution and assume the graph $G$ has spectral decomposition $A = \sum_{r} \theta_r E_r$. Let $\widehat{M}$ be the standard average mixing matrix and $\Delta_{rs} := \theta_r - \theta_s$. Then

\begin{equation}
\label{equationhatM_R}
    \widehat{M}_R = \widehat{M} + 2\sum_{s < r} (E_r \circ E_s)\ \mathbb{E}[\cos{(\Delta_{rs}R)}].
\end{equation}

\begin{proof}
Let all be defined as in the statement, then the following sequence of equalities hold

\begin{align*}
\widehat{M}_R  &= \mathbb{E}[M(R)]\\
&= 
\int_{-\infty}^{\infty}M(t)f_R(t)\dd t  \\
& = \int_{-\infty}^{\infty}U(t) \circ \overline{U(t)}f_R(t)\dd t \\
& = \int_{-\infty}^{\infty}\sum_r\sum_s \e^{\ii\Delta_{rs} t}E_r \circ E_s\ f_R(t) \dd t\\
& = \int_{-\infty}^{\infty}\left(\sum_r E_r \circ E_r + \sum_{r \neq s} \e^{\ii\Delta_{rs} t} E_r \circ E_s \right) f_R(t) \dd t\\
&= \sum_r E_r \circ E_r\left(\int_{-\infty}^{\infty}f_R(t)\dd t\right) + \sum_{r \neq s} E_r \circ E_s \left(\int_{-\infty}^{\infty}\e^{i\Delta_{rs} t}f_R(t)\dd t\right)\\
&= \sum_rE_r \circ E_r + 2 \sum_{r < s} (E_r \circ E_s) \int_{-\infty}^{\infty}\cos{(\Delta_{rs}t)}f_R(t)dt \\
&= \sum_rE_r\circ E_r + 2\sum_{s < r}E_r\circ E_s\  \mathbb{E}[\cos{(\Delta_{rs}R)}],
\end{align*}
where the fourth equality holds because $U(t) = \sum_r \e^{\ii t\theta_r}E_r$ and $ \overline{U(t)} = U(-t)$, the fifth equality because $\Delta_{rr} = 0$ for all $r$, the seventh because $\int_{-\infty}^{\infty}f_G(t)dt = 1$, $E_r \circ E_s = E_s \circ E_r$ and Euler's identity, and the last follows from Proposition~\ref{propsExpecPSF}.
\end{proof}
\end{proposition} 

\subsection{Properties of $\widehat{M}_R$}
We prove now some basic properties of the matrix $\widehat{M}_R$ defined in Definition~\ref{def:amm}. Some of these are similar or analogous to known results for the standard average mixing matrix (see \cite{Godsil2013}).

\begin{proposition}
\label{theoPropertAvgMix}
The average mixing matrix $\widehat{M}_R$ satisfies the following properties.
\begin{enumerate}[(a)]
    \item $\widehat{M}_R$ is symmetric.
    \item $\widehat{M}_R$ is doubly stochastic.
    \item $ I \succeq \widehat{M}_R \succeq 2\widehat{M} - I$.
    \item The eigenvalues of $\widehat{M}_R$ lie in the interval [-1, 1].
    \item $tr(\widehat{M}_R) \ge 2\tr (\widehat{M}) - n$
\end{enumerate}

\begin{proof}
\begin{enumerate}[(a)]
    \item Follows straightforwardly from \eqref{equationAvgMixMat} and \eqref{equationhatM_R} as each $E_r$ is symmetric.
    \item $\widehat{M}_R = \int_{-\infty}^{\infty}M(t)f_R(t)dt$ is an average of mixing matrices, and since $M(t)$ is doubly stochastic, so is $\widehat{M}_R$.
    \item Recall that the $E_r$ are projectors and thus positive semi-definite, then by the Schur Product theorem $E_r \circ E_s \succeq 0$. Also notice that $-1 \le \mathbb{E}[\cos(\Delta_{rs}R)] \le 1$. Finally, note that $\sum_r E_r = I$ implies
    \[
    	I = I \circ I = \sum_r E_r \circ E_r + 2 \sum_{r < s} E_r \circ E_s.
    \]
    Both inequalities then follow immediately.
    \item Consequence of item (c) and $\widehat{M} \succeq 0$.
    \item Follows from item (c).
\end{enumerate}
\end{proof}
\end{proposition}

\section{Uniform average mixing under discrete distributions}\label{discrete}

Under the classical definition of average mixing matrix, the only graph which admits uniform average mixing (Definition~\ref{def:uniform}) is $P_2$, the path on two vertices (see \cite[Lemma 17.2]{GodsilStateTransfer12}). 

This section is motivated by the problem of finding a graph $G$ and a probability distribution $R$ such that $\widehat{M}_R$ is uniform. We now present a necessary condition for that to happen.

\begin{lemma}
\label{LemMultipliInequ}
    Assume $G$ has $n$ vertices and denote the multiplicity of eigenvalue $\theta_r$ by $m_r$. If there is a $R$ such that $\widehat{M}_R = \frac{1}{n}J$, then
    \[
\frac{n+1}{2} \ge \tr (\widehat{M}) \quad \text{and} \quad \tr (\widehat{M}) \geq \frac{1}{n}\sum_r{m_r^2}.
	\]

\begin{proof}
On one hand, if $\widehat{M}_R = \frac{1}{n}J$, then $\tr{(\widehat{M}_R)} = 1$. Thus Proposition~\ref{theoPropertAvgMix} item (e) gives that
\[ \tr (\widehat{M}) \le \frac{n+1}{2}.\]
On the other hand, by recalling that $\tr (\widehat{M}) = \sum_r { \tr(E_r \circ E_r})$ and $m_r = \tr(E_r)$, an easy application of Cauchy-Schwarz 
	gives
	\[
		\tr (\widehat{M}) \geq \frac{1}{n} \sum_r m_r^2.
	\]
\end{proof}    
\end{lemma}
(The second inequation is known \cite[Lemma 4]{CoutinhoGodsilSurvey}).

In the following subsections, we use some of the properties we developed to construct novel examples of uniform average mixing, as well as prove some no-go results.

\subsection{Complete graphs and complete bipartite graphs}
\label{subsectionCompleteGraph}

Complete graphs are graphs in which all vertices are connected. They are fully symmetric, and, for classical random walks, they display very quick mixing behaviour (as long as there are more than two vertices). For quantum walks, however, they tend to contradict a first intuition: for instance, there is no time $t$ for which the entries of $M(t)$ is constant (as long as $n \geq 5$), and in \cite{godsil2021sedentary} it was shown that the average behaviour of a quantum walk in a complete graph is to stay at the initial vertex. Besides for $K_2$, they also have no uniform average mixing. A natural question is: is there a choice of sampling distribution that fixes this? The answer is no.

\begin{theorem}
\label{PropComplGrap_Great5}
For the complete graphs $K_n$, the following hold:
\begin{itemize}
\item With $n \ge 5$, there is no distribution $R$ such that $\widehat{M}_R = \frac{1}{n}J$.
\item With $n \le 4$, there is an $R$ such that $\widehat{M}_R = \frac{1}{n}J$.
\end{itemize}
\begin{proof}
The spectral idempotents of the complete graph $K_n$ are $E_0 = (1/n)J$ and $I-E_0$. Thus,
\[\tr(\widehat{M})= n - 2 + \frac{2}{n}.\]

By Lemma~\ref{LemMultipliInequ}, if there exists $R$ such that $\widehat{M}_R = \frac{1}{n}J$, then 
\[n - 2 + \frac{2}{n} \le \frac{n+1}{2},\] which implies $(n-1)(n-4) \le 0$. Therefore, for complete graphs, $n \le 4$ is a necessary condition for uniform average mixing under $R$.

For $2 \leq n \leq 4$, the complete graph admits instantaneous uniform mixing (see \cite{adamczak2003note}) respectively at times $\pi/4$, $2\pi/9$ and $\pi/4$, that is, $M(t)$ is constant at these times. Thus, by choosing $R$ to be a random variable with density $1$ at these values, we can tweak $\widehat{M}_R$ to be constant. In other words, letting $a_n = \mathbb{E}[\cos(nR)]$, it follows from Proposition~\ref{PropDefinAvgMixMat} that $\widehat{M}_R  = (1/n)J$ if and only if
\[
0 = I\left(1 - \frac{2}{n} + 2a_n\frac{1}{n}\right) + J\left(\frac{2}{n^2}-2a_n\frac{1}{n^2} - \frac{1}{n}\right),
\]
which is equivalent to $a_n = 1 - (n/2)$. For $n=2$, we have $\mathbb{E}[\cos(2 R)] = 0$ if $R$ is constant equal to $\pi/4$, corresponding to $f_R(t) = \delta_{\frac{\pi}{4}}(t)$. For $n=3$, $\mathbb{E}[\cos(3 R)] = -(1/2)$ if $R$ is constant equal to $2\pi/9$, corresponding to $f_R(t) = \delta_{\frac{2\pi}{9}}(t)$. For $n=4$, $\mathbb{E}[\cos(4 R)] = -1$ if $R$ is constant equal to $\pi/4$, corresponding to $f_R(t) = \delta_{\frac{\pi}{4}}(t)$.
\end{proof}
\end{theorem}

The intuition should now indicate that few eigenvalues with high multiplicity are bad indicators of uniform average mixing. We show another class of examples. The complete bipartite graphs $K_{m,n}$ are obtained upon fully joining two sets of $m$ and $n$ vertices, respectively, leaving no edge within each of the sets.

\begin{proposition}
For the complete bipartite graph $K_{m,n}$ with $m+n > 7$, there is no uniform average mixing under any $R$.

\begin{proof}
The spectrum of $K_{m,n}$ is $\pm \sqrt{mn}$ and $0$ with multiplicities, $1$, $1$ and $m+n-2$, respectively. Let $v = m + n$, then by Lemma~\ref{LemMultipliInequ}, in order for uniform average mixing to exist for some $R$, we must have
$$ 1^2 + 1^2 + (v-2)^2 \le \frac{v(v+1)}{2} \implies v^2 -9v + 12 \le 0.$$
This condition can only be satisfied if $v < 8$.
\end{proof}
\end{proposition}

We are left with $10$ possible complete bipartite graphs to verify the existence of an $R$ that presents uniform average mixing. Four cases are known:

\begin{itemize}
    \item $K_{1,1} = P_2$: $\widehat{M}$ is uniform.
    \item $K_{1,2} = P_3$: There exists, Subsection \ref{subsecP3Dirac}.
    \item $K_{1,3}$: There is instantaneous uniform mixing at $\frac{2\pi}{3\sqrt{3}}$ (see \cite{GodsilZhan}), so one can create a Dirac distribution for it, as shown in the proof of Theorem~\ref{PropComplGrap_Great5}.
    \item $K_{2,2} = C_4$: There is instantaneous mixing at $\frac{\pi}{4}$ (see \cite{adamczak2007non}).
\end{itemize}

\subsection{Uniform average mixing on $P_3$}
\label{subsecP3Dirac}

In this section, we display a much more interesting example. We will now show that we can create a discrete-like probability density function that allows uniform average mixing to appear in the path graph on three vertices, denoted by $P_3$ (this might also be known as the linear chain on three nodes). 

\begin{corollary}
	Let $G = P_3$. There is $R$ so that $\widehat{M}_R = (1/3)J$.

\begin{proof}
Let
\[
A(P_3) = \begin{pmatrix}0 & 1 & 0 \\ 1 & 0 & 1 \\ 0 & 1 & 0\end{pmatrix}\]
be the adjacency matrix of $P_3$. Its eigenvalues are $\{\sqrt{2}, \ 0, \ -\sqrt{2}\}$ with respective idempotents 
\[E_{\sqrt{2}} = \frac{1}{4}\begin{pmatrix}1 & \sqrt{2} & 1 \\ \sqrt{2} & 2 & \sqrt{2} \\ 1 & \sqrt{2} & 1\end{pmatrix}, \ \ E_{0} = \frac{1}{2}\begin{pmatrix}1 & 0 & -1 \\ 0 & 0 & 0 \\ -1 & 0 & 1\end{pmatrix}, \ \ E_{-\sqrt{2}} = \frac{1}{4}\begin{pmatrix}1 & -\sqrt{2} & 1 \\ -\sqrt{2} & 2 & -\sqrt{2} \\ 1 & -\sqrt{2} & 1\end{pmatrix}.\]
Let $a_{rs} = \mathbb{E}[\cos(\Delta_{rs}R)]$. From Proposition~\ref{PropDefinAvgMixMat}, in order for uniform average mixing to happen there must exist $a_{12}, a_{13}, a_{23}$ such that:

\begin{align*}
\frac{1}{n} J = \frac{1}{8}\begin{pmatrix} 3 & 2 & 3 \\ 2 & 4 & 2 \\ 3 & 2 & 3\end{pmatrix} + \frac{a_{12}}{4}\begin{pmatrix} 1 & 0 & -1 \\ 0 & 0 & 0 \\ -1 & 0 & 1\end{pmatrix} + \frac{a_{13}}{8}\begin{pmatrix} 1 & -2 & 1 \\-2 & 4 & -2 \\ 1 & -2 & 1\end{pmatrix} + \frac{a_{23}}{4}\begin{pmatrix} 1 & 0 & -1 \\0 & 0 & 0 \\ -1 & 0 & 1\end{pmatrix}.
\end{align*}

The only solution for the constants $a_{12}, a_{13}, a_{23}$ is 
\begin{equation}
\begin{cases}
\label{casesP3Dirac}
    a_{12} = a_{23} = \mathbb{E}[\cos(\sqrt{2}R)] = 0,\\
    a_{13} = \mathbb{E}[\cos(2\sqrt{2}R)] = - 1/3.
\end{cases}
\end{equation}

The problem now is whether there is a probability distribution that satisfies these conditions over its expectations. It is possible to verify (using Proposition~\ref{propsExpecPSF} for instance) that 
\[f_G(t) = \frac{1}{2}(\delta_{\theta}(t) + \delta_{\frac{\pi}{\sqrt{2}} - \theta}(t)), \text{ for } \theta = \frac{\arccos{\frac{-1}{3}}}{2\sqrt{2}},\]
satisfies both conditions. 

\end{proof}
\end{corollary}

Note that the solution for $a_{12}, a_{13}, a_{23}$ is unique, however we will see in Subsection \ref{subsecP3gauss} that there are infinitely many distributions that satisfy the conditions in \eqref{casesP3Dirac}.

\section{$\widehat{M}_R$ and the Characteristic Function}\label{char}

Even though the example in Subsection~\ref{subsecP3Dirac} is interesting, it could be arguably obtained without making use of the machinery from Section~\ref{sec:2}. Our goal in this section is to present a formulation of $\widehat{M}_R$ that helps us work with a wider variety of probability distributions, such as continuous ones.

Associated with a random variable $R$, there is a probability density function $f_R(t)$ that satisfies two conditions 

\[\int_{-\infty}^{\infty}f_R(t) \dd t = 1 \text{ and } f_R(t) \ge 0\ \forall \ t \in \mathbb{R}.\]

Often in probability theory, working directly with the probability density function is a bit difficult. Certain problems require an alternative representation of a random variable, and in our context, it will be particularly useful to consider the \textit{characteristic function} (see \cite[Chapter XV]{feller1991introduction}).

\begin{definition} Let $R$ be a random variable with probability density function $f_R(x)$. Its characteristic function is defined as.

\[\phi_R(w) := \mathbb{E}[\e^{\ii w R}] = \int_{-\infty}^{\infty}\e^{\ii w x}f_R(x)dx .\]
\end{definition}

The characteristic function is closely related to the Fourier transform of the probability density function, and it is often interpreted as such. Therefore, many of the properties of the Fourier transform are reflected by the characteristic functions. We list some of them below for reference (again, check \cite[Chapter XV]{feller1991introduction} for these and for the characteristic function of some common distributions).

\begin{proposition}
\label{propCharacFunc}
$\phi_R(w)$ has the following properties:
    \begin{enumerate}[(i)]
        \item $\phi_R (0) = 1, |\phi_R (w)| \le 1$.
        \item $\phi_R (-w) = \overline{\phi_R (w)}$. \label{itemConjugCharacFunc}
        \item $\phi_{aR + b}(w) = \e^{\ii b}\phi_R (aw)$.
        \item If $R$ and $V$ are two independent random variables, then $\phi_{R + V}(w) = \phi_{R}(w) \phi_{V}(w)$. \label{itemProdCharacIndependVar}
    \end{enumerate}
\end{proposition}

We present $\widehat{M}_R$ in terms of the characteristic function.

\begin{proposition}
\label{PropDefhatM_R_CharacFunc}
Let $R$ be a random variable with characteristic function $\phi_R(w)$ and $\Delta_{rs} := \theta_r - \theta_s$. Then $\widehat{M}_R$ takes the form:
$$\widehat{M}_R = \widehat{M} + 2\sum_{s < r} E_r \circ E_s [\mathtt{Re}(\phi(\Delta_{rs}))]$$
\begin{proof}
We proceed similarly as in the proof of Proposition~\ref{PropDefinAvgMixMat}.
\begin{align*}
    \widehat{M}_R  &= \mathbb{E}[M(R)]\\
    &= \sum_r E_r \circ E_r + \sum_{r \neq s}E_r \circ E_s \int_{-\infty}^{\infty}e^{i\Delta_{rs} t}f_R(t)dt\\
    &= \sum_r E_r \circ E_r + \sum_{r \neq s}E_r \circ E_s\ \mathbb{E}(\e^{\ii\Delta_{rs}R})\\
    &= \sum_r E_r \circ E_r + \sum_{r \neq s}E_r \circ E_s\ \phi(\Delta_{rs})\\
    &= \sum_r E_r \circ E_r + \sum_{s < r}E_r \circ E_s\ [\phi(\Delta_{rs}) + \phi(-\Delta_{rs})] \\
    &= \sum_r E_r \circ E_r + \sum_{s < r}E_r \circ E_s\ [\phi(\Delta_{rs}) + \overline{\phi(\Delta_{rs})}]\\
    &= \sum_r E_r \circ E_r + 2\sum_{s < r}E_r \circ E_s\ [\mathtt{Re}(\phi(\Delta_{rs}))].
\end{align*}
\end{proof}
\end{proposition}

\subsection{$P_3$ uniform average mixing under Gaussian distribution}
\label{subsecP3gauss}

As shown in Subsection~\ref{subsecP3Dirac}, for $P_3$, using Dirac's function, the only solution is the one with coefficients as in the cases shown \eqref{casesP3Dirac}. However, in the characteristic function formulation of $\widehat{M}_R$, $a_{12} = a_{23} = \mathtt{Re}(\phi_R(\sqrt{2}))$ and $a_{13} = \mathtt{Re}(\phi_R(2\sqrt{2}))$. Thus, the system of equations becomes:
\begin{equation}
\label{casesP3Fourier}
\begin{cases}
    \mathtt{Re}(\phi_R(\sqrt{2})) = 0,\\
    \mathtt{Re}(\phi_R(2\sqrt{2})) = - 1/3.
\end{cases}
\end{equation}

Let $R \sim \texttt{Gaussian}(\mu, \sigma^2)$, then its characteristic function and real part have the form:

\[\phi(w) = \exp\left(\ii w\mu - \frac{1}{2}\sigma^2w^2\right)\implies Re(\phi(w)) = \cos(w\mu)\e^{-\frac{1}{2}\sigma^2w^2}\]

\begin{itemize}
\item The first equation in \eqref{casesP3Fourier} requires: 

\begin{equation}
\label{eqCosP3Gaus1}
    \cos(\sqrt{2}\mu) = 0 \implies \mu = \frac{\pi}{2\sqrt{2}} + \frac{k\pi}{\sqrt{2}}; k \in \mathbb{Z}.
\end{equation}

\item Given $\mu$ of the form in \eqref{eqCosP3Gaus1}, the second equation in \eqref{casesP3Fourier} imposes that the value of $\sigma^2$ must be:

\begin{align*}
-\frac{1}{3} &= \cos(2\sqrt{2}\mu)\ \e^{-4\sigma^2} \\ 
&= \cos(\pi + 2k\pi)\ \e^{-4\sigma^2} \\ 
&= -\e^{-4\sigma^2} \implies \sigma^2 = \frac{1}{4}\ln(3).
\end{align*}
\end{itemize}

Therefore, sampling from any Gaussian distribution of the form 
\[R \sim \texttt{Gaussian}\left(\mu = \frac{\pi}{2\sqrt{2}} + \frac{k\pi}{\sqrt{2}};\ \sigma^2 = \frac{1}{4}\ln(3)\right),\]
for $k \in \mathbb{Z}$, results in a uniform $\widehat{M}_R$.

\subsection{$K_3$ with Gaussian and Bernoulli distributions}

To illustrate the usefulness of characteristic functions in this context, we present additional examples of uniform average mixing with continuous distributions.

\begin{example}$K_3$ has average uniform mixing under a Gaussian Sampling.
\end{example} 

In Subsection~\ref{subsectionCompleteGraph}, we showed that $K_3$ has average uniform mixing if $\mathbb{E}(\cos(3R)) = -\frac{1}{2}$, for $R$ coming from a single Dirac function. In terms of a characteristic function, this means $\mathtt{Re}(\phi_R(3)) = -\frac{1}{2}$. Suppose $R \sim \texttt{Gaussian}(\mu; \sigma^2)$, then

$$\mathtt{Re}(\phi_R(3)) = \cos(3\mu)\e^{-\frac{9}{2}\sigma^2} = -\frac{1}{2}.$$

There are infinitely many solutions for $(\mu, \sigma^2)$ in this case. Let $\mu = \frac{\pi}{3} + \frac{2k\pi}{3}$, for $ k \in \mathbb{Z}$. Then

$$\e^{-\frac{9}{2}\sigma^2} = \frac{1}{2} \implies \sigma^2 = \frac{2}{9}\ln(2).$$

Thus, sampling from any Gaussian distribution of the form \[R \sim \texttt{Gaussian}\left(\mu = \frac{\pi}{3} + \frac{2k\pi}{3}; \ \sigma^2 = \frac{2}{9}\ln(2)\right)\] results in an uniform $\widehat{M}_R$.

\begin{example} $K_3$ has uniform mixing under a Bernoulli sampling.
\end{example}

If $R \sim \texttt{Bernoulli}(p)$, then we choose time $1$ with probability $p$ and time $0$ with probability $1-p$. Its characteristic function and real part take the form

$$\phi_R(t) = 1 - p + p \e^{\ii t} \implies \mathtt{Re}(\phi_R(t)) = 1 - p + pcos(t).$$

Solving $\mathtt{Re}(\phi_G(3)) = -\frac{1}{2}$ for $p$ results in

$$p = \frac{3}{2(1-\cos(3))} \approx 0.753.$$

Thus, sampling from a $\texttt{Bernoulli}\left(\frac{3}{2(1-\cos(3))}\right)$ results in an uniform $\widehat{M}_R$.\\

\subsection{$P_4$ uniform average mixing}

In previous sections, we showed that $P_2$ and $P_3$ have uniform average mixing. Generalizing those results for larger paths is not easy. In this subsection, we show that $P_4$ also has uniform average mixing, and we use a slightly different method from the ones above.

The spectrum of $P_n$ is known (see for instance \cite{BrouwerHaemers}). The characteristic polynomial of $P_4$ has degree four, and its roots are the roots of $p(x) = x^2-x-1$, say $\alpha$ and $\beta$, and their negatives. Using the fact that the eigenspaces are invariant under the non-trivial automorphism of the path, it is easy to express the spectral idempotents in terms of $\alpha$ and $\beta$. For instance, in \cite{Godsil2013} a similar reasoning was used to compute the standard average mixing matrix of paths explicitly. In our context, we can derive that $P_4$ has uniform average mixing if there are constants $d_i$ which satisfy
\[
	\frac{1}{4}J = \widehat{M} + 2d_1 E_{\alpha} \circ E_{\beta} + 2d_2 E_{\alpha} \circ E_{-\alpha} + 2d_3 E_{\alpha} \circ E_{-\beta} + 2d_4 E_{\beta} \circ E_{-\alpha} + 2d_5 E_{\beta} \circ E_{-\beta} + 2d_6 E_{-\alpha} \circ E_{-\beta},
\]
and, moreover, the $d_i$ corresponding to $E_r \circ E_s$ is equal to $\mathtt{Re}(\phi(r+s))$ with $\phi$ the characteristic function of some random variable $R$. Recall that $\mathtt{Re}(\phi(r+s)) = \mathtt{Re}(\phi(-(r+s)))$, and note that $E_{\alpha} \circ E_{-\beta} = E_{-\alpha} \circ E_{\beta}$ and $E_{\alpha} \circ E_{\beta} = E_{-\alpha} \circ E_{-\beta}$. Using explicit expressions for some of these matrices and also for $\widehat{M}$, the equation above is equivalent to
\begin{align*}
\frac{1}{4}J &= \frac{1}{10}\begin{bmatrix}
    3 & 2 & 2 & 3\\
     & 3 & 3 & 2\\
     &  & 3 & 2\\
     &  &  & 3
     \end{bmatrix} + 2c_1\ E_{\alpha} \circ E_{-\alpha} + 2c_2\ E_{\beta} \circ E_{-\beta}\ + \\
     &+2c_3\ \frac{1}{10}\begin{bmatrix}
    1 & -1 & -1 & 1\\
     & 1 & 1 & -1\\
     &  & 1 & -1\\
     &  &  & 1\end{bmatrix} + 2c_4\ \frac{1}{10}\begin{bmatrix}
    1 & 1 & -1 & -1\\
     & 1 & -1 & -1\\
     &  & 1 & 1\\
     &  &  & 1
     \end{bmatrix},
\end{align*}
with a trivial solution given by
\[\begin{cases}
    & c_1 = c_2 = c_4 = 0,\\
    & c_3 = -\frac{1}{4}.
\end{cases}\]

Thus, it remains to construct a probability distribution whose characteristic function satisfies

\[\begin{cases}
    & \mathtt{Re}(\phi(2\alpha)) = \mathtt{Re}(\phi(2\beta)) = \mathtt{Re}(\phi(\alpha + \beta)) = 0,\\
    & \mathtt{Re}(\phi(\alpha- \beta)) = -\frac{1}{4}.
\end{cases}\]

Consider the discrete distribution that picks times $a$ and $-a$ with probabilities $\frac{1}{2}$. Then, its probability density function and characteristic function are
\begin{align*}
    f(t) &= \frac{\delta_{-a}(t) + \delta_{a}(t)}{2},\ \text{ and}\\
    \phi(s) &= \cos(as) \implies \mathtt{Re}(\phi(s)) = \cos(as).
\end{align*}

Let $R_1, R_2, R_3, R_4$ be four of the aforementioned distributions on times $\mu_1, \mu_2, \mu_3, \mu_4$, respectively. Define a new random variable $T = R_1 + R_2 + R_3 + R_4$. It is a consequence of Proposition~\ref{propCharacFunc} item~\ref{itemProdCharacIndependVar} that
$$\phi_T(s) = \prod_{i=1}^{4}\phi_{R_i}(s) = \prod_{i=1}^{4}\cos(\mu_i s) = \mathtt{Re}(\phi_T(s)).$$
Therefore, we can choose the $\mu$'s to meet the conditions.
\begin{itemize}
    \item We can choose $\mu_1 = \frac{\pi}{4\alpha}$ to satisfy $\mathtt{Re}(\phi(2\alpha))=\prod_{i=1}^{4}\cos(2\alpha\mu_i )=0$.
    \item Analogously, we can satisfy $\mathtt{Re}(\phi(2\beta)) = \mathtt{Re}(\phi(\alpha + \beta)) = 0$ with ${\mu_2 = \frac{\pi}{4\beta}}$ and ${\mu_3 = \frac{\pi}{2(\alpha+\beta)}}$.
    \item Given the above $\mu_1, \mu_2, \mu_3$, we must try to solve the following equation for $\mu_4$:
    \[ \mathtt{Re}(\phi(\alpha - \beta)) = \prod_{i=1}^{4}\cos((\alpha - \beta )\mu_i )=-\frac{1}{4}.\]
We have
    \[\cos\left(\frac{(\alpha-\beta)\pi}{4\alpha}\right)\cos\left(\frac{(\alpha-\beta)\pi}{4\beta}\right)\cos\left(\frac{(\alpha-\beta)\pi}{2(\alpha+\beta)}\right)\cos\left((\alpha-\beta)\mu_4\right) = -0.25\] 
    \[\implies \cos((\alpha - \beta)\mu_4) = \frac{-0.25}{\cos\left(\frac{(\alpha-\beta)\pi}{4\alpha}\right)\cos\left(\frac{(\alpha-\beta)\pi}{4\beta}\right)\cos\left(\frac{(\alpha-\beta)\pi}{2(\alpha+\beta)}\right)} \approx -0.601.\]

    Since the right-hand side of the equation above is a number in the interval $[-1,1]$, we can solve for $\mu_4$.
\end{itemize}

Therefore, a solution for the $\mu_i$ is given by

$$\begin{cases}
     &\mu_1 = \frac{\pi}{4\alpha} \approx 0.485,\\
    & \mu_2 = \frac{\pi}{4\beta} \approx -1.27,\\
    & \mu_3 = \frac{\pi}{2(\alpha+\beta)} \approx 1.57,\\
    & \mu_4 = \frac{1}{(\alpha - \beta)}\arccos{\frac{-0.25}{\cos{\frac{(\alpha - \beta)\pi}{4\alpha}}\cos{\frac{(\alpha - \beta)\pi}{4\beta}}\cos{\frac{(\alpha - \beta)\pi}{2(\alpha + \beta)}}}} \approx 0.9912.
\end{cases}$$

Since the density function of the sum of independent random variables is the convolution of the individual density functions (see \cite[Chapter V]{feller1991introduction}), we have

$$f_T(t) = (f_{R_1} \ast f_{R_2} \ast f_{R_3}\ast f_{R_4})(t)$$
in which 
\[f_{R_i}(t) = \frac{\delta_{-\mu_i}(t) + \delta_{\mu_i}(t)}{2}.\]

Methods described in this section should be available for larger paths, and we ask the question of whether it is always possible for any path graph $P_n$ to find a sampling distribution $R$ for which $\widehat{M}_R$ is constant.

\section{Cartesian Products} \label{cartesian}

If $G$ and $H$ are graphs on $n$ and $m$ vertices, their Cartesian product $G \Box H$ is the graph with vertex set $V(G) \times V(H)$ and adjacency matrix
$$A(G \square H) = A(G) \otimes I_{m} + I_{n} \otimes A(H).$$
It is immediate to check that $G \Box H$ is isomorphic to $H\Box G$. We start this section by presenting a few known results that should help us understand how the average mixing matrix under $R$ of the Cartesian product $\widehat{M}_R^{G \square H}$ relates to the ones of the individual graphs $\widehat{M}_R^{G}$ and $\widehat{M}_R^{H}$. In this section, denote by $M^G(t)$ the mixing matrix of graph $G$ defined previously in \eqref{equationMof_t}, and similarly for $M^G(R)$, as in Definition~\ref{def:amm}. It is known (see \cite{ChristandlPSTQuantumSpinNet2}) that
\begin{align}
	M^{G \square H}(t) = M^G(t) \otimes M^H(t).
\end{align}
One might be tempted to state that $\widehat{M}_R^{G \square H} = \widehat{M}_R^G \otimes \widehat{M}_R^H$ and that if $\widehat{M}_R$ is uniform for both $G$ and $H$, then so it is for $G \square H$. However, these do not hold in general, because we are not sampling from $M^G(t)$ and $M^H(t)$ independently. Nevertheless, it is possible to say something about the traces of these matrices under certain conditions. Recall that if $R$ and $S$ are two random variables, then the \emph{covariance} between $R$ and $S$ is defined as
$$\Cov[R, S] := \mathbb{E}[(R - \mathbb{E}[R])(S - \mathbb{E}[S])].$$
Basic properties of the covariance are listed below for reference (see \cite{schervish2014probability}).
\begin{proposition}
\label{propCovar}
The covariance has the following properties:
\begin{enumerate}[(i)]
    \item It can be written as $\Cov[R, S] = \mathbb{E}[RS] - \mathbb{E}[R]\cdot \mathbb{E}[S]$.\label{propEquCovSumExpec}
    \item It is bilinear.
    \item It is symmetric: $\Cov[R, S] = \Cov[S, R]$.
    \item $|\Cov[R,S]| \le \sqrt{\Var(R)\Var(S)}$.
    \item $\Cov[R, R] = \Var(R) \ge 0$.
\end{enumerate}
\end{proposition}

We now derive the main result of this section.

\begin{theorem}
\label{theorTracCovAvMixMat}
Let $R$ be a sampling distribution and $G$ and $H$ be graphs with mixing matrices $M^G(t)$ and $M^H(t)$, respectively. Then
$$\tr(\widehat{M}_R^{G \square H}) = \Cov[\tr(M^G(R)), \tr(M^H(R))] + \tr(\widehat{M}_R^G) \tr(\widehat{M}_R^H).$$

\begin{proof}
By definition,
$$\widehat{M}_R^{G \square H} = \mathbb{E}[M^{G \square H}(R)] = \mathbb{E}[M^G(R) \otimes M^H(R)].$$
Let $a$ and $b$ be vertices from $G$ and $H$, respectively. Restricting to the diagonal elements of $\widehat{M}_R^{G \square H}$, we have:
\begin{align*}
    (\widehat{M}_R^{G \square H})_{ab, ab} &= \mathbb{E}[(M^{G}(R))_{a,a}(M^{H}(R))_{b, b}]\\
    &= \Cov[(M^{G}(R))_{a,a},(M^{H}(R))_{b, b}] + \mathbb{E}[(M^{G}(R))_{a,a}] \cdot \mathbb{E}[(M^{H}(R))_{b, b}],
\end{align*}
the last equality coming from Proposition \ref{propCovar} item \ref{propEquCovSumExpec}. Summing over all $a,b$ and using the bilinearity of the covariance, we have

\begin{align*}
    \tr(\widehat{M}_R^{G \square H}) &= \sum_{a}\sum_{b}(\widehat{M}_R^{G \square H})_{ab, ab} \\
    &= \sum_{a}\sum_{b}\Cov[(M^{G}(R))_{a,a},(M^{H}(R))_{b, b}] + \sum_{a}\sum_{b}\mathbb{E}[M^{G}(R)_{a,a}]\mathbb{E}[(M^{H}(R)_{b, b}]\\
    &= \Cov[\sum_a(M^{G}(R))_{a,a},\sum_b(M^{H}(R))_{b, b}] + (\sum_a\mathbb{E}[(M^{G}(R))_{a,a}])\ (\sum_b\mathbb{E}[(M^{H}(R))_{b, b}])\\
    &= \Cov[\tr(M^{G}(R)), \tr(M^{H}(R))] + \tr(\widehat{M}_R^G)\tr(\widehat{M}_R^H).
\end{align*}
\end{proof}
\end{theorem}

This gives us the following corollaries.

\begin{corollary}
\label{CorTracCartProdofX_X}
Let $G$ be a graph and $R$ a sampling distribution. Then

$$\tr(\widehat{M}_R^{G \square G}) \ge [\tr(\widehat{M}_R^G)]^2.$$
\begin{proof}
It follows from Theorem \ref{theorTracCovAvMixMat} by noticing that if $G = H$, then $$\Cov[\tr(M^G(R)), \tr(M^G(R))] = \Var(\tr(M^G(R))) \ge 0.$$
\end{proof}
\end{corollary}

\begin{corollary}
\label{corTracCartProdofX_X}
Let $G$ be a graph on $n$ vertices and $\widehat{M}^{G}$ be its classical average mixing matrix. If there exists a $R$ such that $\widehat{M}_{R}^{G \square G} = \frac{1}{n^2}J_{n^2\times n^2}$, then

\[\tr(\widehat{M}^G)\le \frac{n+1}{2}.\]

\begin{proof}
If $\widehat{M}_R^{G \square G}$ is uniform for some $R$, then $\tr(\widehat{M}_R^{G \square G}) = 1$. However, by Corollary \ref{CorTracCartProdofX_X}, we have

$$ 1 = \sqrt{\tr(\widehat{M}_R^{G \square G})} \ge \tr(\widehat{M}_R^G) \ge 2 \tr(\widehat{M}^G) - n.$$

Thus, 
\[\tr(\widehat{M}^G)\le \frac{n+1}{2}.\]
\end{proof}
\end{corollary}

Corollary~\ref{corTracCartProdofX_X} can be used to show that some graphs do not admit uniform average mixing under any $R$.

\begin{corollary} For $K_5 \square K_5$, there is no $R$ such that $\widehat{M}_R^{K_5 \square K_5}$ is uniform.
\begin{proof}
As shown in Subsection~\ref{subsectionCompleteGraph}, for $K_5$,
\[\tr(\widehat{M}^{K_5}) > \frac{n+1}{2}.\]
The result follows from Corollary~\ref{corTracCartProdofX_X}.
\end{proof}

\end{corollary}

\section{Average States} \label{average}

An alternative way of representing quantum states is by using density matrices. They are a generalization of the more usual state vectors or wave functions. In this representation, a state $D$ is a positive semidefinite matrix with trace 1. If $D$ is the initial state of a continuous-time quantum walk, then the state $D(t)$ at time $t$ is
$$D(t) = U(t)DU(-t).$$

We now present results that relate this representation of quantum states with our previous definition of average mixing matrices. 

\begin{definition}
\label{def:avg}
Let $R$ be a probability distribution. Define the average state under $R$ as
\[\Psi_R(D) := \mathbb{E}[D(R)].\]
\end{definition}
From Proposition~\ref{propsExpecPSF}, it follows that
\begin{align*}
 \mathbb{E}[D(R)] & = \int_{-\infty}^{\infty}D(t)f_R(t) \dd t\\ 
&= \sum_ rE_r D E_r + \sum_{r \neq s} E_r D E_s \int_{-\infty}^{\infty} \e^{\ii \Delta_{rs}t}f_R(t) \dd t\\ 
&= \sum_r E_r D E_r + \sum_{r \neq s} E_rDE_s\ \phi_R(\Delta_{rs}).
\end{align*}

Several properties of average states that hold for the uniform distribution (see \cite{Coutinho2018}) are still true more generally, as we show below.

\begin{proposition}
\label{propLinearMapProp}
The average state $\Psi_R(D)$ has the following properties:
\begin{enumerate}[(a)]
    \item $\tr(\Psi_R(D)) = tr(D)$, i.e., it is trace-preserving.
    \item $\Psi_R(D)$ is positive semi-definite.
    \item $\Psi_R(I) = I$, i.e., it is unital.
\end{enumerate}
In particular, $\Psi_R(D)$ is a quantum state.
\begin{proof}
    \begin{enumerate}[(a)]
    \item \begin{align*} \tr(\Psi_R(D)) &= \sum_r \tr(E_r D E_r) + \sum_{r \neq s} \tr(E_r D E_s)\ \phi_R(\Delta_{rs})\\
    &= \sum_r \tr(D E_r^2) + \sum_{r \neq s} \tr(D E_r E_s)\ \phi_R(\Delta_{rs})\\
    &= \sum_r \tr(DE_r) \\ & = \tr(D).
    \end{align*}
    
    \item Since $D(t)$ is positive semi-definite (and bounded) and $f_R(t) \ge 0\ \forall \ t \in \mathbb{R}^+$, then $\int_{-\infty}^{\infty}D(t)f_R(t)\dd t$ is also positive semi-definite matrix.
    \item Follows easily: $\Psi_R(I) = \sum_r E_r I E_r + \sum_{r \neq s} E_r I E_s\ \phi_G(\Delta_{rs})
    = \sum_r E_r^2
    = I.$
\end{enumerate}
\end{proof}
\end{proposition}

Is $\Psi_R$ a quantum channel? A linear map that is unital, trace-preserving and positive semi-definite is a good candidate to be a quantum channel. It remains to show that it is completely positive. So the question is: for any $k \in \mathbb{Z}_+$, is it the case that $\text{id}_k \otimes \Psi_R$ maps positive semi-definite matrices to positive semi-definite matrices? 

The characterization of completely positive maps in \cite[Theorem 1]{Choi1975} implies that if $R$ is the uniform distribution on the real line (as in \cite{Coutinho2018}), then $\Psi_R$ is a quantum channel. Fortunately, the set of unital completely positive maps is a convex set (see \cite{Choi1975} and \cite{Landau1993}). Thus, just as in Proposition~\ref{propLinearMapProp} item~(b), one concludes that $\Psi_R$ is a quantum channel. 

Before presenting the next result, it is important to note the following:

\begin{proposition}
\label{diffProb}
Let $R_1, R_2$ be two independent random variables and let $\phi_{R_1}(w),\   \phi_{R_2}(w)$ be their characteristic functions. The characteristic function of $R_2 - R_1$ is given by

$$\phi_{R_2 - R_1}(s) = \phi_{R_2}(s)\overline{\phi_{R_1}(s)}$$

\begin{proof} Noting that $R_1, R_2$ are independent and from Proposition \ref{propCharacFunc} items \ref{itemConjugCharacFunc} and \ref{itemProdCharacIndependVar}, we have

\begin{align*}
\phi_{R_2 - R_1}(w) &= \phi_{R_2}(w)\phi_{R_1}(-w) = \phi_{R_2}(w) \overline{\phi_{R_1}(w)}
\end{align*}
\end{proof}    
\end{proposition}

Let $G$ be a graph, then the quantum state representing $a \in V(G)$ is given by $D_a = e_a e_a^T$, in which $e_a$ denotes the standard basis vector of $\mathbb{C}^{V(G)}$ (alternatively, we could denote ${D_a = \ket a \bra a}$.) The following theorem connects the averages of these states and the average mixing matrices.

\begin{theorem}
\label{avstateAMM}
Let G be a graph and $R_1, R_2$ be two independent random variables with associated average state maps $\Psi_{R_1}$ and $\Psi_{R_2}$ (as in Definition~\ref{def:avg}). Then 

$$(\widehat{M}_{R_2 - R_1})_{a,b} = \langle \Psi_{R_1}(D_a),\Psi_{R_2}(D_b) \rangle,\ \forall\ a, b \in V(G).$$

\begin{proof}
First, note that
\begin{align*}
\Psi_{R_1}(D_a)^*\Psi_{R_2}(D_b)  &= \sum_r E_r D_a E_r D_b E_r + \sum_{r\neq q}E_r D_a E_r D_b E_q\ \phi_{R_2}(\Delta_{rq})+\\
&+\sum_{s \neq r}E_s D_a E_r D_b E_r\ \overline{\phi_{R_1}(\Delta_{rs})}+\sum_{s \neq r \neq q} E_s D_a E_r D_b E_q\ \overline{\phi_{R_1}(\Delta_{rs})} \phi_{R_2}(\Delta_{rq}).
\end{align*}
Whence

\begin{align*}
 \langle \Psi_{R_1}(D_a),\Psi_{R_2}(D_b) \rangle &= \tr(\Psi_{R_1}(D_a)^{*}\Psi_{R_2}(D_b))\\
&=\sum_r \tr(E_r D_a E_r D_b E_r)+ \sum_{s\neq r}\tr(E_s D_a E_r D_b E_s)\ \overline{\phi_{R_1}(\Delta_{rs})}\phi_{R_2}(\Delta_{rs})\\
&= \sum_r \tr(D_a E_r D_b E_r) + \sum_{s \neq r}\tr(D_a E_r D_b E_s)\ \phi_{R_2 - R_1}(\Delta_{rs})\\
&= \sum_r \tr(e_a^T E_r e_b e_b^T E_r e_a) + \sum_{s \neq r}\tr(e_a^T E_r e_b e_b^T E_s e_a)\ \phi_{R_2 - R_1}(\Delta_{rs})\\
&= \sum_r (E_r)_{a,b} (E_r)_{b,a} + \sum_{r \neq s} (E_r)_{a,b} (E_s)_{b,a} \ \phi_{R_2 - R_1}(\Delta_{rs}) \\
&= \sum_r (E_r^{\circ 2})_{a,b} + \sum_{r \neq s} (E_r \circ E_s)_{a,b}\  \phi_{R_2 - R_1}(\Delta_{rs})\\
&= (\widehat{M}_{{R_2 - R_1}})_{a,b}
\end{align*}
\end{proof}
\end{theorem}

This allows us to conclude a corollary analogous to \cite[Theorem 5]{Coutinho2018}.

\begin{corollary} Let $G$ be a graph and be $R_1, R_2$ two independent and identically distributed random variables. Then $\widehat{M}_{R_2 - R_1}$ is the Gram Matrix of the average states $\Psi_{R_1}(D_a)$ for $a \in V(G)$.

\begin{proof} Follows from Theorem \ref{avstateAMM} noting that, since $R_1,\ R_2$ are identically distributed, then $\Psi_{R_1}(D_a) = \Psi_{R_2}(D_a)\ \forall\ a \in V(G)$.
\end{proof}    
\end{corollary}

\section{Open questions}

We have shown in Sections~\ref{discrete} and \ref{char} how to find alternative sampling distributions to achieve a desired expected effect on a quantum walk. Our most relevant examples were achieved by exhibiting unexpected behaviour on the paths on three and four vertices. This motivates the question
\begin{itemize}
	\item Is there always a sampling distribution that allows for uniform average mixing on $P_n$? We suspect the answer to this question to be positive, but we prefer to state no conjecture.
\end{itemize}
Several properties related to combinatorics and symmetries of graphs were obtained in \cite{Coutinho2018} by exploring the fact that the (standard) average mixing matrix is the Gram matrix of certain average states. We managed to show an analogous result, but we have not yet found some meaningful combinatorial consequences. We believe that analytically finding the Kraus decomposition for the quantum channel $\Psi_R$ (at least in some particular cases) might lead to interesting consequences.

Finally, we wonder if there is a nice connection between the technology we develop in this paper and other quantum walk phenomena, such as perfect or pretty good state transfer.

\section*{Acknowledgements}

Vitor Marques acknowledges the PICME scholarship. Pedro Baptista acknowledges the support from CNPq. Gabriel Coutinho acknowledges the support from CNPq and FAPEMIG. 

\bibliographystyle{unsrt}
\bibliography{qbib.bib,pbib.bib}

\end{document}